\documentclass[english,11pt,aps,prd,a4paper,preprintnumbers,floatfix,nofootinbib,showpacs,superscriptaddress, notitlepage]{revtex4-1} 
\usepackage[mode=buildnew]{standalone}
\usepackage[usenames,dvipsnames]{color}  
\usepackage{graphicx}

\usepackage{setspace}
\usepackage{caption}
\usepackage{amsmath}
\usepackage{amssymb}
\usepackage[colorlinks=true,citecolor=darkred,urlcolor=darkred, pdfborder={0 0 0}]{hyperref}
\usepackage[normalem]{ulem}
\usepackage{xcolor}
\usepackage{placeins}
\usepackage{mathrsfs}
\usepackage{verbatim} 
\usepackage{cancel} 
\usepackage{float} 
\usepackage{multirow}
\usepackage{scalerel}
\usepackage[compat=1.1.0]{tikz-feynhand}
\usepackage{tikz-feynman}
\tikzfeynmanset{compat=1.1.0}
\usepackage{feynmp}
\usepackage{tikzsymbols}
\usepackage{array}
\usepackage{pifont}
\usepackage{subfigure}

\makeatletter
\def\p@subsection{}
\makeatother

\definecolor{darkred}{rgb}{0.6,0,0}

\definecolor{linkcolor}{rgb}{0,0,0.5}

\def\gsim{\raise0.3ex\hbox{$\;>$\kern-0.75em\raise-1.1ex\hbox{$\sim\;$}}}
\def\lsim{\raise0.3ex\hbox{$\;<$\kern-0.75em\raise-1.1ex\hbox{$\sim\;$}}}

\def\beqn#1{\begin{equation}\label{#1}}
\def\eeqn{\end{equation}}

\def\beqa#1{\begin{eqnarray}\label{#1}}
\def\eeqa{\end{eqnarray}}

\newcommand {\ignore}[1]{}

\def\UBL{$U(1)_{B-L}$}
\def\Z2{$Z_2$}
\def\Z3{$Z_3$}
\def\Z4{$Z_4$}
\def\O5{$\mathcal{O}_5$}

\def\321{$\mathrm{SU(3) \otimes SU(2) \otimes U(1)}$ }

\usepackage{orcidlink}
 
 \newcommand{\AddrIISERB}{Department of Physics,
 Indian Institute of Science Education and Research - Bhopal \\
 Bhopal Bypass Road, Bhauri, Bhopal, India}

\begin{document}

\title{\color{BrickRed}
Ultra-Heavy Dark Matter Through Exponential Production Mechanism}

\author{Disha Bhatia\orcidlink{0000-0001-8758-8799}
}\email{disha.bhatia@apu.edu.in}
\affiliation{School of Sciences, Azim Premji University,
Sarjapura, Bangalore, 562125, India}
\author{Ranjeet Kumar\orcidlink{0000-0002-7144-7606}
}\email{kumarranjeet.drk@gmail.com}
\affiliation{Institute for Convergence of Basic Studies, Seoul National University of Science and Technology, Seoul 01811, Republic of Korea}
\affiliation{\AddrIISERB}
\author{Rahul Srivastava\orcidlink{0000-0001-7023-5727}
}\email{rahul@iiserb.ac.in}
\affiliation{\AddrIISERB}

\begin{abstract}
\vspace{1cm} 
\noindent
Exponential production of dark matter has been introduced recently as one of the alternatives to the widely studied thermal production of dark matter scenarios. 
This non-thermal production mechanism leads to the exponential growth in the abundance of the dark matter ($\chi$) through its scatterings 
with the bath particle ($\xi$), viz. $\chi + \xi \to \chi + \chi$. 
In our framework, we consider a scalar dark matter $\chi$ charged under $Z_3$ symmetry and bath particle $\xi$ to be standard model singlet
and determine the maximum mass of the dark matter using the criteria of perturbativity. 
We find that dark matter has an upper mass around the ballpark of $\mathcal{O}(10^{15})$ GeV for the cases when the bath particle mass is similar to the dark matter mass. The computations are largely performed with minimal model dependence and will hold in any scenario where the renormalizable couplings driving
the exponential growth effectively reduce to $\chi^3\xi$ interactions. Thus, in exponential production
of dark matter, we have a possibility of an ultra-heavy dark matter with large couplings, making
the scenario phenomenologically interesting.
Later, we extend the simplified dark matter model to a UV-complete Dirac Type-I seesaw model where the $Z_3$ symmetry is obtained 
as a residual unbroken subgroup of a global $U(1)_{B-L}$ symmetry, providing a natural connection with the neutrino sector.

\end{abstract}
\maketitle

\section{Introduction}
\label{sec:intro}

Despite compelling astrophysical and cosmological evidence for dark matter (DM) \cite{Zwicky:1933gu,Rubin:1970zza,Rubin:1980zd,Planck:2018vyg}, its nature and production mechanism remain among the most pressing questions in the field of particle and astroparticle physics~\cite{Lisanti:2016jxe}. The mass of dark matter is allowed to span a remarkably broad range, from as low as $\mathcal{O}$($10^{-21}$) eV to as high as $\mathcal{O}(10^{3}M_\odot)$, based on model-independent considerations~\cite{PhysRevLett.134.151001,Carr:1997cn}. There are various mechanisms which predict the correct relic abundance of dark matter in accordance with the cosmic microwave radiation data~\cite{Planck:2018vyg}. Different production mechanisms point toward different ranges of {\it dark matter mass}, which can then be probed by various types of experiments depending on feasibility.

Among the dark matter production mechanisms, thermal freeze-out is one of the most studied mechanisms~\cite{Scherrer:1985zt}. In this scenario, dark matter particles were initially in thermal equilibrium with either the Standard Model (SM) bath or a hidden sector thermal bath in the early Universe. As the Universe expanded and cooled, the annihilation rate of dark matter into the bath particles eventually fell below the Hubble expansion rate, causing the dark matter particles to depart from chemical equilibrium. Their annihilations subsequently became inefficient, creating a finite relic abundance of dark matter particles. Within the category of thermally produced dark matter scenarios, weakly interacting massive particles (WIMPs)~\cite{Lee:1977ua,Kolb:1990vq,Jungman:1995df} have been the most studied because of the weak-scale interactions of the dark matter, which can be probed easily at terrestrial experiments. 

\begin{figure}[h!]
    \centering  
     \includegraphics[width=0.45\textwidth]{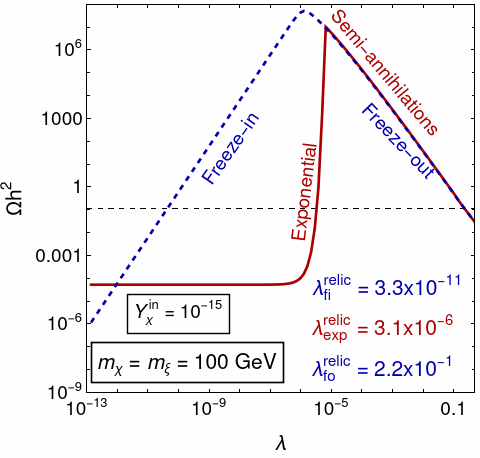}
    \caption{The phase diagram for various dark matter production mechanisms is shown for $m_\xi =m_{\chi} = 100$ GeV. The $m_\chi$ and $m_\xi$ are the masses of the dark matter candidate $\chi$ and the thermal bath particle $\xi$. The $\lambda^{\rm relic}_i$; $i = {\rm fi}, {\rm fo}, {\rm exp}$ values correspond to the coupling values for which the correct relic density is obtained in each scenario. The numerical values mentioned here are using the simplified Lagrangian given in Appendix~\ref{app:finout}.}
    \label{fig:phasedigintro}
\end{figure}

Besides thermal mechanisms, dark matter in the early universe may be produced via non-thermal mechanisms 
which comprise a wide range of possibilities, viz., freeze-in, exponential production, gravitational production of dark matter, primordial black holes (PBHs), or axions as dark matter~\cite{Cirelli:2024ssz}. 
In this work, we focus on a particular class of non-thermal scenarios in which the thermal bath plays an important role in the production of dark matter particles. 
Freeze-in~\cite{Hall:2009bx} and exponential growth mechanisms~\cite{Bringmann:2021tjr,Hryczuk:2021qtz}
fall under the category of thermally assisted dark matter production mechanisms where the dark matter is not in equilibrium with the thermal bath but is produced either due to scatterings or decay of the bath particles. These mechanisms are consistent with experimental bounds largely because of the feeble couplings between dark matter and the SM particles~\cite{XENON:2023cxc,PandaX:2024qfu,LZ:2024zvo}. 
In Fig.~\ref{fig:phasedigintro}, we show the ballpark couplings required to generate the correct dark matter abundance for certain thermal and non-thermal mechanisms through a phase diagram.

As can be seen from the plot, for the same mass of the dark sector particles, the coupling values needed to obtain the correct relic abundance in the freeze-out scenario are much larger than those needed in the exponential growth case, which in turn is larger than that needed in the freeze-in scenario\footnote{Note that the numerical analysis in Fig.~\ref{fig:phasedigintro} has been computed using the simple renormalizable model discussed in detail in Appendix~\ref{app:finout}, however the results are obtained are generic $\lambda_{\rm fo} \gg \lambda_{\rm exp} \gg \lambda_{\rm fi}$}. 

In this work, we focus on the recently introduced exponential growth production mechanism of dark matter~\cite{Bringmann:2021tjr,Hryczuk:2021qtz,Bhatia:2023yux},
where the scatterings between dark matter particle $\chi$ and the bath particle $\xi$, i.e., $\chi \ + \ \xi \to \ \chi + \ \chi$, 
increase the dark matter abundance exponentially. This mechanism is inherently initial condition dependent as it requires the presence 
of a small initial abundance of dark matter to kickstart the exponential growth\footnote{{The initial abundance can be generated via any 
non-thermal process~\cite{Cirelli:2024ssz}}.}.
The exponential growth of dark matter naturally stops once the process becomes Boltzmann suppressed.

The earlier works on the exponential production mechanism have focused on the ideas and complementarity with the semi-annihilation mechanism~\cite{Bringmann:2021tjr,Hryczuk:2021qtz}, applicability in the sterile neutrino sector~\cite{Bringmann:2022aim} and determination of the initial condition dependence on the phase space distribution 
functions~\cite{Bhatia:2023yux}. However, a dedicated analysis focusing on the determination of the upper bound of dark matter in these scenarios was not performed.
This maybe perhaps due to the fact that generally non-thermal scenarios allow for a wide range 
of masses. This is certainly true for freeze-in produced dark matter, the upper bound here sensitively depends on the kind of freeze-in model considered, i.e., whether freeze-in is IR dominated or UV dominated, the mass of the mediator, reheating temperature, etc.~\cite{Bernal:2017kxu,Cirelli:2024ssz}. However, on the other hand, for freeze-out scenarios, the model-independent upper bounds exist from perturbative unitarity~\cite{Griest:1989wd,Hui:2001wy,Bhatia:2020itt,Gaurav:2026esz}.
It may seem apparent that perhaps a wide range of masses may be allowed for the exponentially produced dark matter scenarios, as it is also a non-thermal mechanism.

 In our framework, we analyze the question of the upper bound on the exponential production of dark matter. 
We perform the analysis using a simplified model of a scalar dark matter ($\chi$) charged under $Z_3$ symmetry and interacting with a Standard Model singlet bath particle $\xi$. 
Like freeze-in, the upper bound of dark matter gets sensitively related to the model parameters -- the initial dark matter comoving densities, 
the mass ratios between dark matter and the thermal bath particle $\xi$, and the interaction strength. In this work, we answer the following question: how heavy the dark matter can be in exponentially produced scenarios. We determine this answer using perturbativity. 
We find that the approximate upper bound on dark matter in this scenario is around the ballpark of the Grand Unification scale, i.e., $\mathcal{O}(10^{15})$ GeV.  This limit is obtained for the case when the bath particle and dark matter mass are comparable. The upper mass bounds reduce when the bath particle is lighter than the dark matter owing to compensation due to Boltzmann suppression of the cross-section.
Our analysis is performed with minimal model dependence, and the broad results will hold for similar classes of renormalizable models.


Moreover, we also extend our simplified model to a UV-complete model. The extension is made along the lines of addressing both neutrino mass generation and exponential growth of dark matter together.
We utilize the accidental $U(1)_{B-L}$ symmetry and its residual $Z_m$ subgroup
to connect neutrino mass generation with exponential growth of dark matter. 
The possibility of realizing either Dirac or Majorana neutrinos from the breaking of $U(1)_{B-L}$ to a residual $Z_m$ symmetry has been explored in Refs.~\cite{Ma:2014qra,Ma:2015mjd,Bonilla:2018ynb,CentellesChulia:2019gic}.
The minimal choice, $Z_2$, allows Majorana neutrinos but does not accommodate exponential dark matter growth scenario.
We therefore consider $Z_3$, which allows exponential growth and neutrinos are Dirac in nature, and present a UV-complete model framework. More generally, exponential growth of dark matter can be realized for $Z_{3n}$ symmetries, allowing both Dirac and Majorana neutrinos.
This framework not only offers a unified origin for exponential dark matter growth and neutrino mass generation but also highlights the role of residual discrete symmetries~\cite{Kumar:2024zfb,Kumar:2025cte,Kumar:2025aek,Kumar:2025zvv} in connecting non-thermal dark matter production with neutrino physics.


The paper is organized as follows. In Sec.~\ref{sec:dm}, we describe our simplified model of dark matter and general properties of the exponential production mechanism. 
In Sec.~\ref{sec:uppermass}, we discuss the upper mass limit of dark matter in exponentially produced scenarios. In Sec.~\ref{sec:model}, we discuss the UV-complete models relating 
the dark matter to the neutrino sector, and comment on the complementarity of this connection. Finally, in Sec.~\ref{sec:conc}, we summarize and present our concluding remarks.

\section{Exponential production of Dark Matter in the Early Universe} 
\label{sec:dm}
As the name suggests, in the exponential production of dark matter, the initial dark 
matter abundance exponentially increases via its interactions with the thermal bath particles. 
This mechanism was recently proposed as a complementary non-thermal mechanism to semi-annihilation~\cite{Bringmann:2021tjr,Hryczuk:2021qtz}.
The abundance of dark matter grows until the scattering process $\chi + \xi \to \chi + \chi$ becomes 
Boltzmann suppressed. This can happen in two scenarios: when the bath particle is heavier than $\chi$, 
this happens when $\xi$ becomes non-relativistic and decays to other bath particles. In the opposite 
case, where dark matter is heavier than the bath particle, the process becomes Boltzmann suppressed when 
there is not enough energy available in the initial states to create two dark matter particles. 

In this section, we will focus on the generalized discussion on the exponential production mechanism, and discuss the scenarios where this will be the dominant mode of production. For this, we adopt a 
simplified framework of a scalar dark matter particle interacting with another scalar bath particle~\cite{Bringmann:2021tjr,Hryczuk:2021qtz,Bhatia:2023yux}.
To make exponential production the dominant production mechanism, we assign the following charges to 
the bath and dark matter particles: $\chi$ is charged as $\omega$ (where $\omega^3 = 1$ is cube root of unity), and $\xi$ is a singlet (i.e. $\omega^0 = 1$) under $Z_3$ symmetry. 
With the above charges, the following dark matter interaction term can be written,\footnote{Note that the full set of quartic Lagrangian terms allowed by the $Z_3$ symmetry are given in Eq. \eqref{eqn:intLag} of  Appendix~\ref{app:finout}. In addition to the term in Eq. \eqref{eqn:simp}, it also contains the self-interaction terms of the fields and their interaction with Higgs.}  
\begin{eqnarray}
    \mathcal{L}_{\rm simp} = \frac{\lambda_{\rm exp}}{3!} \xi \chi^3  + \text{h.c.} \;.
    \label{eqn:simp} 
\end{eqnarray}
Note that the $Z_3$ charge assignment $\chi \to \omega \chi$, with $\omega = e^{2\pi i/3}$ and $\xi \to \xi$, 
is the minimal choice consistent with the cubic interaction $\xi \chi^3$ responsible for exponential production. This also protects $\chi$ from decaying into bath particles alone, ensuring its stability on cosmological timescales. In Sec.~\ref{sec:model}, we will discuss how this $Z_3$ symmetry will appear naturally due to spontaneous breaking of $U(1)_{B-L}$ symmetry.

\begin{figure}[!h] 
\centering
      \scalebox{1.05}{\begin{tikzpicture}
\begin{feynman}
\vertex (a1) ;
\vertex (a2) ;
\vertex (a3) ;
\vertex (i1) ;

\vertex[right=3.0cm of i1] (i2);
\vertex[right=1.5cm of i1] (a2);
\vertex[right=3.0cm of a1] (a5);
\vertex[right=3.0cm of a3] (a6);
\vertex[below=1.5cm of i1] (a1) {\(\xi \)};
\vertex[above=1.5cm of i1] (a3) {\(\chi \)};
\vertex[below=1.5cm of i2] (a5){\( \chi\)};
\vertex[above=1.5cm of i2] (a6){\( \chi\)};
\diagram* {
(a1) -- [scalar,black,very thick] (a2),
(a3) -- [scalar,black, very thick] (a2),
(a2) -- [scalar,very thick] (a6), 
(a2) -- [scalar, very thick ] (a5), 
};
\end{feynman}
\end{tikzpicture}
\vspace{0cm}
\hspace{1.5cm}
\begin{tikzpicture}
\begin{feynman}
\vertex (a1) ;
\vertex (a2) ;
\vertex (a3) ;
\vertex (i1)  {\(\xi \)};

\vertex[right=3.25cm of i1] (i2)  {\(\chi \)};
\vertex[right=1.5cm of i1] (a2);
\vertex[right=3.0cm of a1] (a5);
\vertex[right=3.0cm of a3] (a6);
\vertex[below=1.5cm of i1] (a1);
\vertex[above=1.5cm of i1] (a3) ;
\vertex[below=1.5cm of i2] (a5){\( \chi\)};
\vertex[above=1.5cm of i2] (a6){\( \chi\)};
\diagram* {
(i1) -- [scalar,black,very thick] (i2),
(a2) -- [scalar,very thick] (a6), 
(a2) -- [scalar, very thick ] (a5), 
};
\end{feynman}
\end{tikzpicture}
\vspace{1cm}
\hspace{1.5cm}} 
    \caption{A schematic of Feynman diagrams showing the scattering and decay processes arising from the $\xi \chi^3$ interaction. Note that we have not explicitly labeled the particles as $\chi$ or $\chi^\dagger$.}
    \label{fig:exp}
\end{figure}
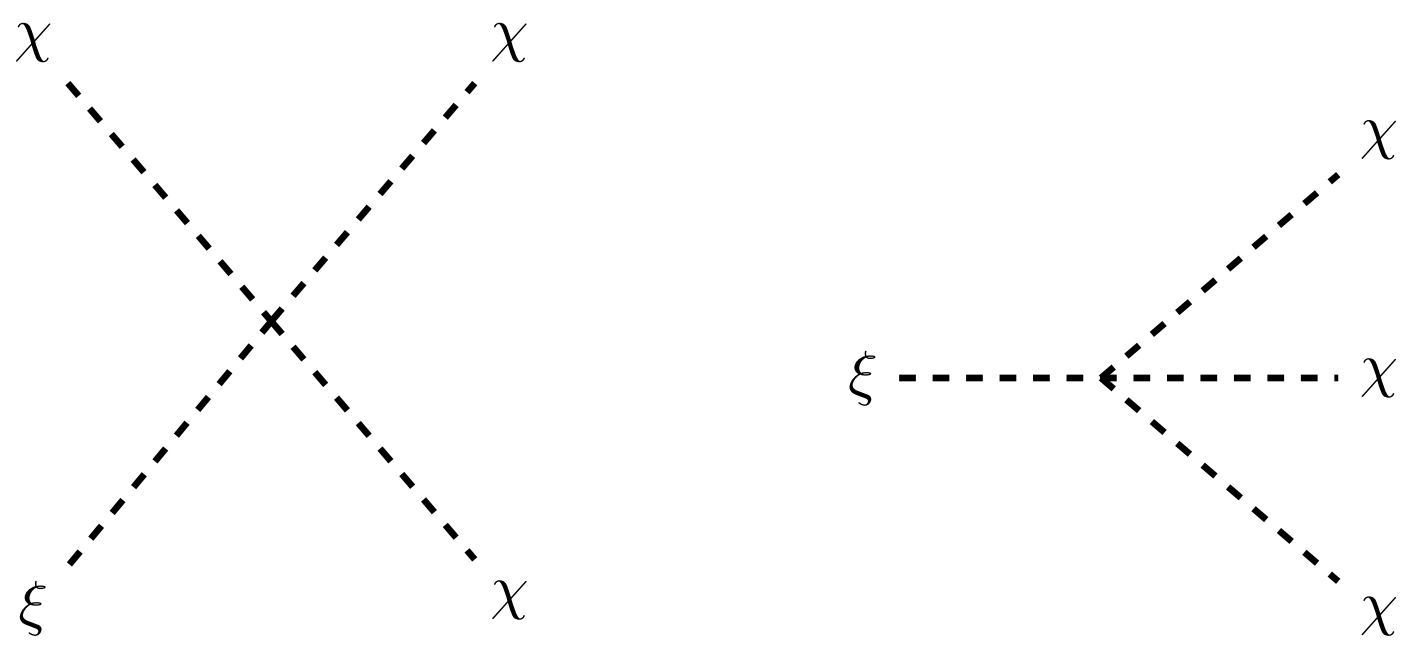

Note that the interactions in Eq.~\eqref{eqn:simp} will lead to dark matter production via the following scatterings:
$\chi + \xi \to \chi^\dagger + \chi^\dagger$ and $\chi^\dagger + \xi \to \chi + \chi$. These interactions, in addition, will
also produce dark matter via direct decay of $\xi \to \chi^\dagger \chi^\dagger \chi^\dagger$, if kinematically 
allowed. The corresponding Feynman diagrams are shown in Fig. \ref{fig:exp}.
For the exponential decay to be the dark matter's dominant production method, we require: 
\begin{enumerate}
    \item Some small initial dark matter population in the early universe to kickstart the exponential growth. 
    \item Mass of the bath particle to be smaller than $3 m_\chi$ such that the decay  $\xi \to \chi^\dagger \chi^\dagger \chi^\dagger$ is kinematically forbidden.
\end{enumerate}

To determine the dark matter abundance generated via the exponential production mechanism, one has to solve the unintegrated Boltzmann equation to determine 
the distribution function $f_\chi(p,t)$, as this is 
apriori a non-thermal production. This is a numerically time taking process~\cite{Hryczuk:2021qtz,Bhatia:2023yux} and the result may potentially 
change depending on how the initial population of dark matter was generated.
However, it was shown in reference~\cite{Bhatia:2023yux} that as the growth in exponential scenarios is rapid,
the lower comoving momentum modes trace the equilibrium behavior. These modes come to equilibrium faster owing to their low mean 
free paths, making the final distribution function for large parts of the comoving momentum initial condition independent. 
Hence, to a reasonable approximation, to get the broad behavior, it is 
simpler to approximate the distribution function of dark matter, in the case of exponential production mechanism, as a scaled equilibrium distribution function, i.e., 
\begin{eqnarray}
    f_\chi(p,t) \approx A f_\chi^{\rm eq}(p,t) \;. 
    \label{eqn:scaled}
\end{eqnarray}

With this approximation, we can directly solve the zeroth-moment Boltzmann equation to determine the dark matter relic density.
For scenarios with no asymmetry between particle and anti-particle
dark matter states, i.e., the case where the chemical potentials of $\chi$ and $\xi$ are zero,
the zeroth-moment Boltzmann equation for dark matter number density $n_\chi$ is given by
\begin{equation}
  \frac{1}{a^3} \frac{d}{dt}\left[ n_\chi(t) a^3 \right] = \langle \sigma v \rangle_{\rm exp} \left( n_\xi n_\chi - \frac{n_\xi}{n_\chi^{\rm eq}}n_\chi^2 \right).
\end{equation}
Here $n_\xi$ is the equilibrium number density for the bath particle, $a$ is the scale factor, and $\langle \sigma v \rangle_{\rm exp}$\footnote{The thermally averaged cross section is given as
\begin{eqnarray}
    \langle \sigma v \rangle_{\rm exp}(x) &=& \frac{2\pi^2 T (2\pi)^6}{n_\chi^{\rm eq}(T) n_\xi^{\rm eq}(T) S}
    \int_{\left( m_\chi + m_\xi\right)^2}^{\infty} \frac{\sigma_{\rm exp}}{\sqrt{s}} \left( s - (m_\xi-m_\chi)^2\right)
    \left( s - (m_\chi+m_\xi)^2 \right) K_1{\left(\frac{\sqrt{s}}{T}\right)} \; \text{where}\;, \nonumber \\
\sigma_{\rm exp} &=& \frac{|\lambda|_{\rm exp}^2 }{16 \pi \sqrt{s}} \frac{\sqrt{s-4 m_\chi^2}}{    \left(s-\left(m_\xi-m_\chi\right)^2\right)^{1/2}  \left(s-\left(m_\xi + m_\chi\right)^2\right)^{1/2}} \;.
\label{eqn:avg}
\end{eqnarray} 
Note that the integration limit is different if the dark matter is heavier than the bath particle, in that case, the integral starts from $4 m_\chi^2$
instead of $\left( m_\chi + m_\xi\right)^2$.} is the thermal
averaged cross-section for the scattering $\xi + \chi \to \chi + \chi$.
Since this is a non-thermal process, the backward process, where dark matter scatters to produce the bath particle, is suppressed.
Hence, we can safely neglect the back-scattering terms. We can also re-express the time variable in terms of the dimensionless parameter $x = m_\xi/T$ (where $m_\xi$ is the mass of the bath particle interacting with dark matter and $T$ is the temperature of the thermal bath)
and the dark matter number density in terms of the comoving number density, i.e., $Y_\chi(x) = n_\chi(x)/s(x)$, where
$s(x)$ is the entropy density of the thermal bath. The zeroth-moment Boltzmann equation
then approximates to
\begin{equation}
  \frac{d Y_\chi(x)}{d x} \approx \frac{h_{\rm eff}(x)\, s(x)}{x\, H(x)}
  \langle \sigma v\rangle_{\rm exp}\, Y_\chi(x)\, Y_\xi^{\rm eq}(x) \;.
  \label{eqn:onlyexp}
\end{equation}
The comoving relic abundance is then
\begin{equation}
  Y_\chi(x) = Y_\chi(x_{\rm init})\,
  \exp\!\left( \int_{x_{\rm init}}^{x}
  \frac{h_{\rm eff}(x')\, s(x')}{x'\, H(x')}\,
  \langle \sigma v\rangle_{\rm exp}(x')\, Y_\xi^{\rm eq}(x')\, dx' \right).
  \label{eqn:onlyexp_sol}
\end{equation}
It can be seen that the dark matter abundance in this case grows exponentially.
The final abundance is sensitive to the initial abundance $Y_\chi(x_{\rm init})$,
and it is exponentially sensitive to the coupling strength $\lambda_{\rm exp}$. 
Hence the dark matter abundance for a given mass can be adjusted by appropriately varying the initial 
abundance and the coupling strength. The dark matter abundance also depends on the mass of the 
bath particle and the dark matter particle as in Eq.~\eqref{eqn:avg}. Note that the mass of the bath particle, for 
this mechanism to be the dominant mechanism for dark matter production, must be smaller than three times the mass of the 
dark matter, i.e., $m_\xi < 3 m_\chi$. 
However, the bath particle can be slightly 
heavier than the dark matter as long as it satisfies the above criteria.
The bath particle
can also be lighter than the dark matter particle. 
The minimum mass of particles in equilibrium with the SM bath from Big Bang Nucleosynthesis is 
approximately 10 MeV~\cite{bbn}. Hence, from the requirement of $m_\xi < 3 m_\chi$, the minimum mass of 
dark matter is roughly $10/3$ MeV. This effectively forms the lower limit for the mass of such a dark matter candidate.
In the following sub-sections, 
we determine the dependence of the dark matter on the model parameters -- $m_\xi$, $m_\chi$, $\lambda_{\rm exp}$
and on the initial abundance $Y_\chi(x_{\rm init})$. 

\subsection{Dependence of dark matter abundance on the model parameters}
The simplified model in Eq.~(\ref{eqn:simp}) has three model parameters: the masses of the dark matter $m_\chi$ and the bath particle $m_\xi$, 
and the coupling between them $\lambda_{\rm exp}$. 
These parameters enter the thermally averaged cross-section (Eq.~\ref{eqn:avg}), and eventually 
affect the net abundance.  In this section, we numerically determine the dependence of 
dark matter relic density on the model parameters. For our analysis, we consider the following
 benchmark scenario: $m_\xi=m_\chi$. We also fix the dark matter initial comoving abundance as $Y_{\chi}(x_{\rm init}) \equiv Y_\chi^{\rm in} = 10^{-15}$. 
Using these, we solve the full Boltzmann equation,\footnote{We solve the full Boltzmann equation, which means we also consider the back-scattering term. This is because we are interested in getting the broad features of this mechanism.} i.e.,
\begin{equation}
  \frac{d Y_\chi(x)}{d x} = \frac{h_{\rm eff}(x)\, s(x)}{x\, H(x)}
  \langle \sigma v\rangle_{\rm exp}\, \left( Y_\chi(x)\, Y_\xi(x) -  \frac{Y_\xi(x)}{Y_\chi^{\rm eq}} Y_\chi^2(x) \right) \;,
  \label{eqn:onlyexp}
\end{equation}
in {\tt Mathematica} \cite{Mathematica}. The analysis is similar to one shown in Fig.~\ref{fig:phasedigintro} in the introduction. However this 
has been performed for wide ranges of the dark matter mass. 

\begin{figure}
    \centering
    \includegraphics[width=0.6\linewidth]{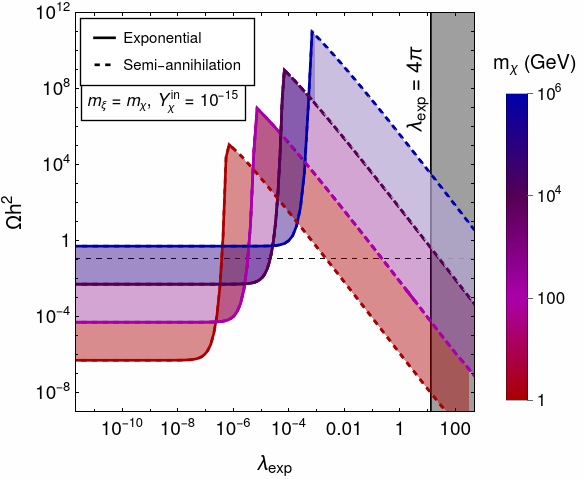}
    \caption{The phase diagram for dark matter abundance as a function of coupling $\lambda_{\rm exp}$ for the benchmark $m_\xi = m_\chi$. Dark matter's initial comoving density is chosen to be $Y_{\chi}(x_{\rm init}) \equiv Y_\chi^{\rm in} = 10^{-15}$ for $x_{\rm init}=0.1$. The figure has been plotted for multiple values of dark matter mass ranging from 1 GeV to $10^6$ GeV. It can be seen that for very small couplings, there is no change in dark matter's abundance, followed by the exponential growth for intermediate couplings and a transition to semi-annihilations for $\mathcal{O}$(1) couplings. The figure also includes a gray band at the coupling $\lambda_{\rm exp} = 4 \pi$ representing the perturbative limit on the interaction strength between dark matter and the thermal bath particle $\xi$.}
    \label{fig:phasediag3}
\end{figure}

In Fig.~\ref{fig:phasediag3}, we plot our numerically obtained results. We observe several interesting features from the plot. We list them here:
\begin{enumerate}
    \item The complementarity between exponential production and semi-annihilations is clearly visible from the plot. Since we vary the 
    coupling over large parts of parameter space, we can see that the correct relic abundance is satisfied for two values: a feeble coupling and 
    $\mathcal{O}(0.01-1)$ coupling depending on the mass. The feeble coupling corresponds to the case of exponential production, and the larger 
    coupling corresponds to the case where dark matter is produced through semi-annihilations. 
    \item In the regions where exponential production is dominant, the abundance grows as  $e^{\lambda_{\rm exp}^2}$, leading to a rapid enhancement, while in the semi-annihilation regime it decreases as $1/\lambda_{\rm exp}^2$.
    \item For couplings which are extremely small, when neither exponential production is dominant nor semi-annihilations are active, the final comoving density of dark matter stays the same as the initial comoving density, i.e., $Y_\chi(x=x_{\rm fin}) = Y_\chi(x=x_{\rm init})$. The dark matter abundance in this case is proportional to $m_\chi \times Y_\chi(x=x_{\rm init})$. Therefore, for a fixed initial comoving density of dark matter, the final abundance would scale with mass. For the same comoving density, the heavier dark matter will be more abundant. This feature can be clearly seen in Fig.~\ref{fig:phasediag3}. Hence, a sufficiently heavy dark matter can attain its correct relic abundance without undergoing any additional scatterings in the later universe. We term this as the no-scattering production region.

    \item  For the regions where the exponential production mechanism is active, as we increase  $m_\chi$, the coupling required for the correct relic abundance also increases. Hence, the heavier dark matter may exhibit interesting phenomenology owing to larger couplings. This feature is in contrast with the IR-dominated freeze-in models, where the coupling required to obtain the correct relic abundance does not depend on the mass of the dark matter particle~\cite{Bernal:2017kxu} and remains feeble for heavier dark matter masses.
\end{enumerate}

In this section, we conclude that, for our chosen benchmark, the dark matter becomes over-abundant for mass scales of approximately $10^6$ GeV. This is true for regions 
where it is produced dominantly through the exponential growth mechanism. We note that the qualitative results obtained in this section remain the same even when $m_\xi \neq m_\chi$ as discussed in Sec.~\ref{sec:uppermass}.

\subsection{Initial condition dependence on the dark matter abundance}
\begin{figure}[htb!]
    \centering
\includegraphics[width=0.55\textwidth]{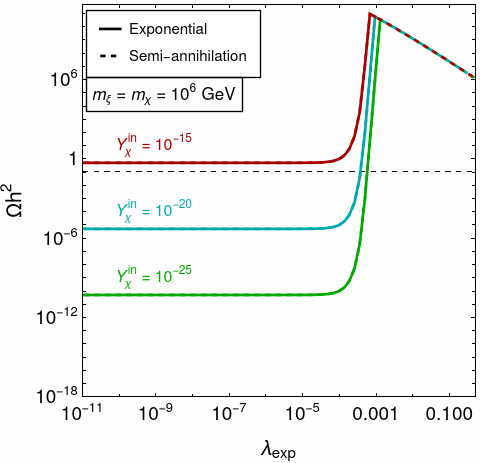}
    \caption{The figure plotted in the plane of $\left( \Omega h^2,\lambda_{\rm exp} \right)$ shows the variation 
    of dark matter relic density with $\lambda_{\rm exp}$ for three different values of the initial abundances i.e. $Y_\chi^{\rm in} = 10^{-15}, 10^{-20}$ and $10^{-25}$. As in Fig.~\ref{fig:phasediag3}, the analysis has been done for the scenario $m_\xi = m_\chi$, which is taken to be $10^6$ GeV. 
     Similar to Fig.~\ref{fig:phasediag3}, we can see regions of no-scattering, exponential growth, and semi-annihilations.}
    \label{fig:rel_33}
\end{figure}
 Using the same benchmark as in the previous section, i.e., $m_\xi = m_\chi$, we determine the dependence of dark matter relic density
on its initial abundance. We discussed in the previous Sec.~\ref{sec:dm} that the exponential production mechanism works if there is some small 
dark matter abundance created already in the early universe. We also saw the final dark matter abundance scaled linearly with the initial abundance 
for the cases where back-scattering was ignored (see Eq.~\ref{eqn:onlyexp_sol}). Hence, the scenarios with over-abundant dark matter 
can be tuned to have the correct abundance by lowering the initial dark matter abundance.
We numerically test this by solving Eq.~(\ref{eqn:onlyexp}) for dark matter mass $10^6$ GeV, which 
in the previous section was over-abundant for $Y_\chi(x_{\rm init})=10^{-15}$.  We solve now for three tentative values of the 
initial dark matter abundance, viz., $10^{-15}$, $10^{-20}$ and $10^{-25}$. 
The results are plotted in Fig.~\ref{fig:rel_33}. 

The analysis in Fig.~\ref{fig:rel_33} is done by varying the coupling $\lambda_{\rm exp}$ in the range [$10^{-11}:0.1$]. Hence, as before, we 
see regions of no-scattering, exponential production, and semi-annihilations as dominant production modes for different regions of $\lambda_{\exp}$. 
Specifically, no-scattering happens when the coupling $\lambda_{\exp}$ is much smaller; exponential production occurs for intermediate values of $\lambda_{\rm exp}$ 
and semi-annihilations for $\mathcal{O}(1)$ values of the coupling parameter.
From the figure, we can see that indeed decreasing the initial dark matter comoving density decreases the final abundance for the coupling regions
where non-thermal production of dark matter dominates, i.e., no-scattering and exponential production regions. 
However, for the couplings where dark matter
is in thermal equilibrium with the Standard Model bath and the abundance is set by semi-annihilations, the final abundance is 
independent of the initial abundance --- a hallmark of the thermally produced scenarios. 
To conclude, from Fig.~\ref{fig:rel_33}, we can see that the over-abundant dark matter can be made to have the correct relic abundance for non-thermal scenarios
by changing the initial conditions. The interesting observation from Fig.~\ref{fig:rel_33} can be seen: both initial abundances $Y_\chi^{\rm in} = 10^{-20}$ and $10^{-25}$ can generate the correct relic density of dark matter today via the exponential production mechanism. However, in comparison to $Y_\chi^{\rm in} = 10^{-20}$, for $Y_\chi^{\rm in} = 10^{-25}$, a slightly larger $\lambda_{\rm exp}$ is required to set the correct abundance lower is the initial abundance. This finding can be understood simply by the reason that for cases with slightly lower dark matter abundances, larger interactions of dark matter with the thermal bath are required to reach the observed dark matter relic density today.

To summarize, the coupling required to attain the correct dark matter abundance increases with: 
\begin{itemize}
    \item increasing the dark matter mass (Fig.~\ref{fig:phasediag3}) 
    \item decreasing the initial comoving number density of the dark matter (Fig.~\ref{fig:rel_33}).
\end{itemize}
Therefore, it is an interesting question to ask that, given the exponential production mechanism, 
how heavy can dark matter be? This is because beyond a certain mass and initial comoving abundance, 
the coupling $\lambda_{\rm exp}$ required to get correct relic abundance will violate perturbative/unitary bounds. 
We explore this question in the next Sec.~\ref{sec:uppermass}.

\section{Dark matter upper mass bound in the exponential growth scenarios}
\label{sec:uppermass}
The question regarding the dark matter upper mass bound has been examined in detail for the thermally produced dark 
matter scenarios for both $2 \to 2$~\cite{Griest:1989wd,Hu:2000ke}
and generalized number changing processes~\cite{Bhatia:2020itt}. The bounds were determined using the constraints from perturbative unitarity. 
For the freeze-out case, 
the upper bound is of the order of 100 TeV for dark matter annihilations proceeding via $2 \to 2$ processes. 
A stronger bound on the mass of the dark matter can be attained by the requirement of perturbative coupling\footnote{Note that the two requirements --- perturbative unitarity and perturbativity of the couplings are slightly different. The former bound comes from demanding the S-matrix to be unitary and the latter bound reflects the coupling values where the higher order diagrams can be ignored. The unitary bounds are largely model independent, while perturbative bounds intrinsically depend on the model under consideration.}~\cite{Cirelli:2024ssz,Gaurav:2026esz}.
In our Fig.~\ref{fig:phasediag3}, we impose the requirement of tree level perturbativity and require the coupling ($\lambda_{\rm exp}$) to be smaller 
than $4 \pi$.

In the case of non-thermal processes like freeze-in, 
usually there are no strong constraints on the mass of dark matter. 
This is because most of the 
couplings are much smaller than $\mathcal{O}(1)$ couplings (see Fig.~\ref{fig:phasedigintro}), 
and the parameters can often be adjusted to yield the correct abundance for a wide range of dark matter masses. 
For example, in the case of IR freeze-in, abundance does not depend on 
the mass of the dark matter particle~\cite{Bernal:2017kxu}. In this scenario, the dark matter mass can be heavy with feeble couplings.

The case of exponential production is a bit different. Although the couplings required to establish the 
correct relic abundance are small for masses around 1 GeV (see Fig.~\ref{fig:phasediag3}), they increase
rapidly with increasing mass (Fig.~{\ref{fig:phasediag3}}) and decreasing initial abundance (Fig.~{\ref{fig:rel_33}}) as discussed in the previous section. 
Therefore, in this section, we address the question
of the upper bound on dark matter in the exponential production scenario. 
For the purpose of our analysis, we use the criteria of perturbativity to determine the upper bounds.

\begin{figure}[htb!]
    \centering
    \includegraphics[width=0.55\linewidth]{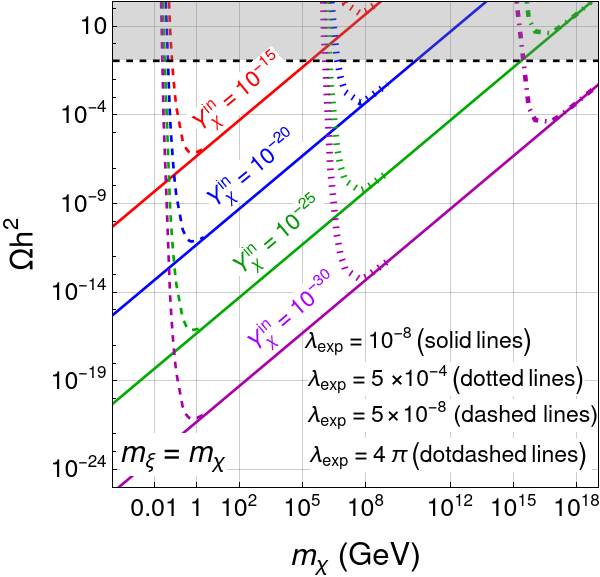}
    \caption{The figure is plotted in the plane of ($\Omega h^2$, $m_\chi$) for a few benchmark values of couplings and dark matter's initial comoving number densities. The solid lines correspond to the cases where no scattering occurs between the dark matter and bath particles. This is numerically implemented by considering a negligibly small coupling of 10$^{-8}$. The dashed/dotted lines correspond to the case where exponential production occurs. The benchmark couplings chosen are $5 \times 10^{-8}$ (dashed), $5 \times 10^{-4}$ (dotted) and $4 \pi$ (dot dashed). The region shaded in gray is disallowed due to overproduction of dark matter. Red, Blue, Green and Magenta colors correspond to $Y_\chi^{\rm in} = 10^{-15}$, $10^{-20}$, $10^{-25}$ and $10^{-30}$ respectively. For this plot, we choose the dark matter and bath particle to have the same mass, i.e., $m_\xi = m_\chi$. }
    \label{fig:mass-limit}
\end{figure}

We begin our analysis for the benchmark case $m_\xi = m_\chi$. Our numerical results are plotted in Fig.~\ref{fig:mass-limit}. To get the mass bounds, we plot the relic abundance of dark matter as a function of dark matter mass for a few selected values of $\lambda_{\rm exp}$. The plot here only highlights the no-scattering and exponential production regions from Fig.~\ref{fig:phasediag3}. We do not show the case of semi-annihilations in this plot.
The solid lines in Fig.~\ref{fig:mass-limit} display the region where no scattering between the dark matter and the bath particle occurs. They correspond to $\lambda_{\rm exp} = 10^{-8}$, which is too small to kickstart the exponential production for the chosen mass range. Different color of these solid lines corresponds to different values of the initial comoving abundances.  
Red, blue, green and magenta lines correspond to initial dark matter comoving number densities as $10^{-15}$, $10^{-20}$, $10^{-25}$ and $10^{-30}$, respectively. In the no-scattering scenario, the final dark matter abundance scales linearly with the dark matter mass as $m_\chi Y_\chi(x=x_{\rm init})$.

The dashed/dotted lines in Fig.~\ref{fig:mass-limit}
correspond to the case where the abundance of dark matter happens via exponential growth. 
As before, red, blue, green, and 
magenta lines correspond to initial dark matter comoving number densities as $10^{-15}$, $10^{-20}$, $10^{-25}$ and $10^{-30}$, respectively. 
Here $\lambda_{\rm exp}$ must be larger than $10^{-8}$ to have exponential production. 
As a representation, we have chosen the following couplings: $5 \times 10^{-8}$ (dashed lines),  $5 \times 10^{-4}$ (dotted lines) and $4\pi$ (dot-dashed lines). 
Since the thermal averaged cross-section is proportional to $\lambda_{\rm exp}^2/m_\chi^2$, for larger masses the dashed/dotted lines merge with the solid lines. This is because the exponential production mechanism 
for larger masses becomes inactive, corresponding to a given $\lambda_{\rm exp}$ and merges into the no-scattering regime. This can be seen explicitly from Fig.~\ref{fig:mass-limit}.

For the dashed/dotted lines in Fig.~\ref{fig:mass-limit}, we can see that upon increasing the coupling values, the masses which set the correct relic abundance also increase and at some mass scale it will hit the perturbative limit. 
We also note that the over-abundant scenarios can be made to have the correct abundance 
by reducing in the initial dark matter abundance. For example, for $\lambda_{\rm exp} = 5 \times 10^{-4}$, the dark matter is over-abundant for $Y_\chi(x_{\rm init})=10^{-15}$, however, acquires the correct abundance for lower values of initial abundance. Note that lowering the initial abundance further for a given coupling reduces the mass of dark matter required for the correct relic density. This shows that for exponential production, in order to get an upper bound on the mass of dark matter from perturbativity, we cannot lower the initial abundance arbitrarily. 
We see that the coupling $4\pi$ points to the mass regions around $10^{15}$ GeV, i.e., around the scales of Grand Unification. 
Hence, for the $4\pi$ coupling, we get roughly the upper bound on the mass of dark matter to be around GUT scales. This result is interesting because it not only accidentally points to the 
Grand Unification scales, which is well motivated by arguments of unification of the three fundamental forces, and it also shows that a scenario which is 
apriori non-thermal can also have upper bounds on masses.

 \begin{figure}[htb!]
     \centering
      \includegraphics[width=0.55\linewidth]{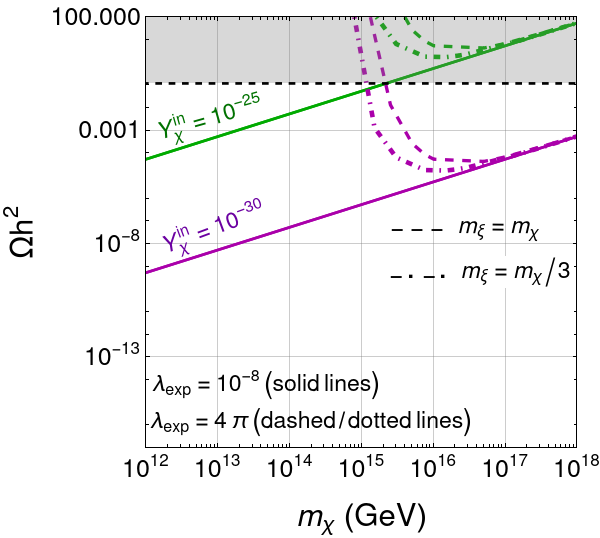}
     \caption{ This is same as Fig.~\ref{fig:mass-limit} but is plotted for different ranges of $m_\xi$ and $m_\chi$ with specific focus on the coupling value $\lambda_{\rm exp} = 4\pi$ shown in dashed/dot-dashed lines. For solid lines the coupling $\lambda_{\rm exp}=10^{-8}$ as before. The initial comoving density of dark matter is chosen to be $Y_\chi(x_{\rm init}) = 10^{-25}$ (in the green line) and $10^{-30}$ (in the magenta line). We represent exponential growth of dark matter for two scenarios: $m_\xi = m_\chi$ (in dashed lines) and $m_\xi = m_\chi/3$ (in dot-dashed lines).}
     \label{fig:ratio-limit}
 \end{figure} 
The result in Fig.~\ref{fig:mass-limit} is valid for a special case, when the dark matter and the bath particle have equal masses. We may ask whether the GUT scale in general is the upper limit for the scenarios when $m_\xi \neq m_\chi$.    
The corresponding results are shown in Fig.~\ref{fig:ratio-limit}. The solid lines in the figure, as before, represent $\lambda_{\rm exp}=10^{-8}$. The dashed/dot-dashed lines correspond to the exponential production and, 
in this case, are plotted for the maximum coupling allowed by perturbativity, i.e., $4 \pi$. The dot-dashed lines in particular correspond to the case where $m_\xi= m_\chi/3$, the dashed line corresponds to $m_\xi = m_\chi$.   
We can see that the maximum mass for dark matter is obtained for the case when $m_\xi$ is around $m_\chi$ and is of the order of $10^{15}$ GeV. 
For the case when the bath particle becomes lighter than the dark matter, the scattering $\xi+\chi \to \chi+\chi$ becomes Boltzmann suppressed, thereby reducing the cross-section. In this case, to get the correct relic abundance for the same coupling, the mass of dark matter is smaller than in the case where scattering rates are not suppressed, this can be seen in Fig.~\ref{fig:ratio-limit} for the case when $m_\xi = m_\chi/3$. 
Thus, the upper limit of dark matter mass in exponential production depends on whether the bath particle is heavier or lighter than the dark matter particle. As discussed before, in order for exponential production to be the dominant production mechanism, the bath particle cannot be much heavier than dark matter to avoid dark matter production by the decays of $\xi$ in particular, $m_\xi < 3 m_\chi$. Hence the bath particle is either of the same order as or smaller than the dark matter mass.


\section{Connecting Exponential Growth of Dark Matter with Neutrino Mass Generation} 
\label{sec:model}

In this section, we discuss the connection between the exponential growth of dark matter and neutrino mass generation. The key interaction responsible for the exponential growth is $\xi \chi^3$, where $\xi$ is a thermal bath particle and $\chi$ serves as the dark matter candidate. This bath particle $\xi$ can also participate in the mechanism responsible for neutrino mass generation. Therefore, these two seemingly disjoint sectors can be connected through the bath particle $\xi$. It then serves as a good motivation to look for models beyond the simplified framework discussed in Sec.~\ref{sec:dm}.
To realize this connection, we utilize the global $U(1)_{B-L}$ symmetry inherent to the SM. The field $\xi$ carries a non-zero $(B-L)$ charge, and its VEV spontaneously breaks the symmetry to its residual $Z_m$ subgroup~\cite{Ma:2014qra,Ma:2015mjd,Bonilla:2018ynb,CentellesChulia:2019gic}. The VEV of $\chi$ can also contribute to the generation of neutrino masses. Moreover, the breaking of $U(1)_{B-L}$ leaves a residual $Z_m$ symmetry, whose structure determines whether neutrinos can be Majorana or Dirac particles \cite{Hirsch:2017col, Bonilla:2018ynb}. To establish a connection with the exponential growth mechanism, we consider the residual symmetry $Z_{3n}$. For odd $n$, this symmetry permits only Dirac neutrinos, whereas for even $n$, both Dirac and Majorana neutrino masses can be realized~\cite{Hirsch:2017col,Bonilla:2018ynb}. The simplest choice is $n=1$, for which the residual symmetry is $Z_3$, allowing Dirac neutrinos and the interaction term $\xi\chi^3$.

\subsection*{Dirac Neutrinos}

We now present a model framework in which the exponential production of dark matter is intimately connected to the Dirac nature of neutrinos and the origin of their masses. 
To generate the neutrino masses, we introduce three right handed neutrinos $\nu_{R_i}$ and three heavy beyond SM Dirac fermion pairs $(N_{L_i},N_{R_i})$ ($i=1,2,3$). Requiring the $U(1)_{B-L}$ symmetry to be anomaly-free fixes their charge assignments to $(-4,-4,5)$ for the $\nu_{R_i}$ and $-1$ for both $N_{L_i}$ and $N_{R_i}$~\cite{Ma:2014qra,Ma:2015mjd,Bonilla:2018ynb,Prajapati:2024wuu,Kumar:2026rul}. The SM leptons $L_i$ and $e_{R_i}$ carry the usual $(B-L)$ charge of $-1$.
The scalar sector is extended by two SM singlet scalars, $\xi$ and $\chi$, in addition to the SM Higgs doublet $\Phi$. The $(B-L)$ charges of scalars ($\Phi, \xi, \chi$) are assigned as ($0,3,1$), respectively.
Their distinct roles are central to the framework: while $\Phi$ and $\xi$ participate in the generation of Dirac neutrino masses through a type-I Dirac seesaw mechanism~\cite{CentellesChulia:2016rms,CentellesChulia:2018gwr}, $\chi$ is stabilized by the underlying symmetry and serves as the dark matter candidate. This field content thus provides the necessary ingredients to realize the common origin of neutrino mass generation and dark matter production within the $U(1)_{B-L}$ framework.
The particle content of the model and its transformation properties under various symmetries are summarized in Tab.~\ref{tab:particle}.
\begin{table}[!h]
\begin{center}
\begin{tabular}{|c| c || c | c | c |c|}
  \hline
& \ Fields \ & \ $ SU(2)_L \otimes U(1)_Y$ \ & \ $U(1)_{B-L}$ $\to$ $Z_3$  \ & \ $Z_2$ \  \\
\hline \hline
  \ \multirow{6}{*}{\rotatebox{90}{Fermions}} \ &$L_{i}$ \ \   & $( 2, -\frac{1}{2})$\ \  &  $-1$ $\to$ \ $\omega^2$ & $1$  \\
&$e_{R_i}$  \ \  & $(1, -1)$  \ \   &  $-1$   $\to$ \ $\omega^2$  & $1$   \\
&$\nu_{R_k}$ \ \   & $( 1, 0)$ \ \  &  $-4$  $\to$ \ $\omega^2$  & $1$ \\
&$\nu_{R_3}$ \ \   & $( 1, 0)$ \ \  & \quad $5$  $\to$ \ $\omega^2$ & $-1$ \\
&$N_{L_i}$ \ \   & $( 1, 0)$ \ \  &  $-1$ $\to$ \ $\omega^2$ & $1$ \\
&$N_{R_i}$ \ \   & $( 1, 0)$ \ \  &  $-1$  $\to$ \ $\omega^2$ & $1$ \\
\hline \hline
\ \multirow{3}{*}{\rotatebox{90}{Scalars}} \ &$\Phi$\ \    & $( 2, \frac{1}{2})$  \ \   &  $0$  $\to$ \ $1$ & $1$ \\
&$\xi$ \ \   & $( 1, 0)$ \ \  &  $3$ $\to$ \ $1$ & $1$ \\
&$\chi$ \ \   & $( 1, 0)$ \ \  &  $1$  $\to$ \ $\omega$ & $1$ \\    
\hline 
  \end{tabular}
\end{center}
\caption{Particle contents and their transformations under different symmetries, where $i=1,2,3$, $k=1,2$, and $\omega$ is a cube root of unity, $\omega^3=1$. The \UBL \ symmetry is spontaneously broken down to its $Z_3$ subgroup by the VEV of the scalar $\xi$. The additional $Z_2$ symmetry forbids the decay term $\bar{\nu}_k^c \nu_{R_3} \chi^\dagger$.}
  \label{tab:particle}
\end{table}

The VEV of the scalar $\langle \xi \rangle = \frac{u}{\sqrt{2}}$ spontaneously breaks the $U(1)_{B-L}$ symmetry down to a residual $Z_3$ symmetry. This residual symmetry plays a crucial role in this framework: it forbids Majorana mass terms, thereby ensuring that neutrinos remain purely Dirac in nature. Simultaneously, it naturally enables an exponential production mechanism that accounts for the observed dark matter relic abundance as discussed in Sec.~\ref{sec:dm}. The scalar $\chi$ does not acquire a VEV, ensuring its stability and making it a viable dark matter candidate. In addition, we impose a discrete $Z_2$ symmetry to forbid the decay term $\bar{\nu}_k^c \nu_{R_3} \chi^\dagger$, under which $\nu_{R_3}$ is odd while all other particles of the model are even.
The \UBL\ symmetry, along with $Z_2$ symmetry, prevents the dark matter candidate $\chi$ from decaying into SM particles. Meanwhile, the residual $Z_3$ symmetry facilitates its exponential growth through the interaction term $\xi^\dagger \chi \chi \chi$, as we have discussed in Sec.~\ref{sec:dm}.

We now discuss the origin of neutrino masses in our framework. Based on the charge assignments listed in Tab.~\ref{tab:particle}, the relevant Yukawa Lagrangian of the leptonic sector is given as follows
\begin{align} \label{eq:Yuklag}
    -\mathcal{L} = y_{ij} \ \overline{L}_{i} \Phi e_{R_j} + f_{ij}\ \overline{L}_{i} \tilde{\Phi} N_{R_j} + g_{i'k}\ \overline{N}_{L_{i'}} \xi \nu_{R_k} + M_{i'j}\ \overline{N}_{L_{i'}} N_{R_j} + \text{H.c.}, 
\end{align}
where $y_{ij}$, $f_{ij}$, and $g_{i'k}$ are Yukawa couplings with indices $i,j,i'=1,2,3$, $k=1,2$. The term $M_{i'j}$ is the matrix element of the mass matrix for the heavy Dirac fermion pairs ($N_{L_{i'}},  N_{R_j}$). 
As a consequence of the underlying \UBL \ charge assignment and its breaking to a residual $Z_3$, Majorana mass terms for neutrinos are forbidden.
In addition, the $\nu_{R_3}$, with a $(B-L)$ charge of $5$, remains decoupled, resulting in one massless neutrino.

\begin{figure}[!h] 
\centering
      \scalebox{1.25}{\input{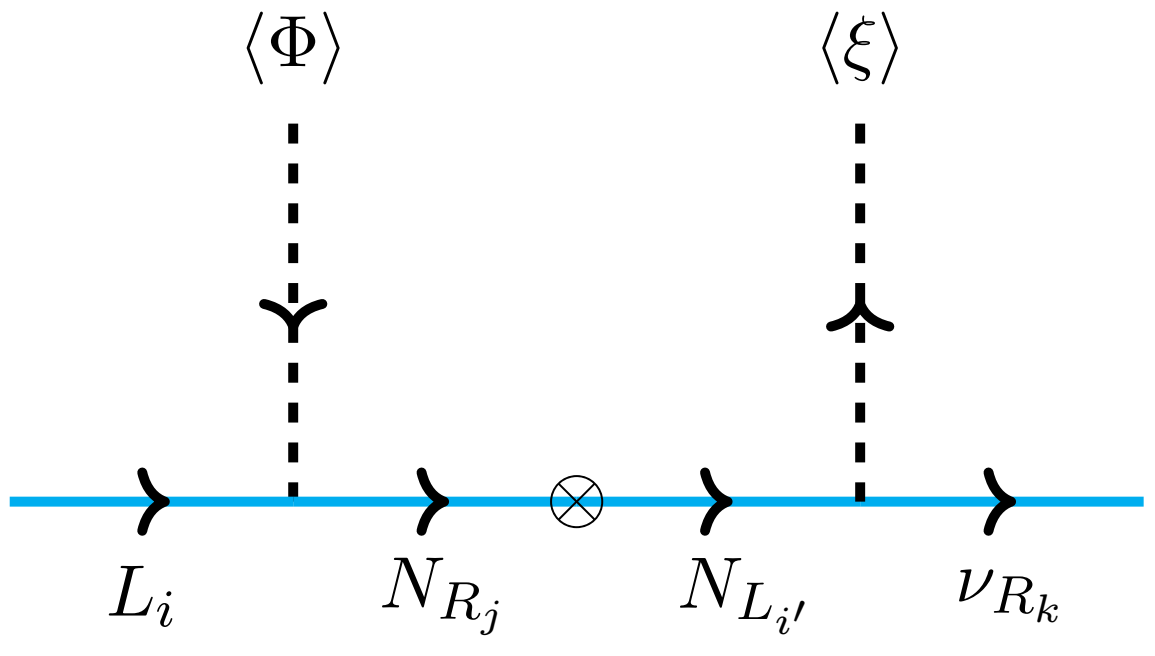}} 
    \caption{Neutrino mass generation through a Dirac type-I seesaw mass mechanism, where $i,j,i'=1,2,3$ and $k=1,2$. Due to the underlying charge assignment, the right handed neutrino $\nu_{R_3}$ remains decoupled.}
    \label{fig:type-i}
\end{figure}

Upon symmetry breaking, the scalars $\Phi$ and $\xi$ acquire vevs, i.e., $\Phi = v/\sqrt{2}$, $\xi = u/\sqrt{2}$, and generate masses for charged lepton and neutrinos.
The neutrino masses are generated through a Dirac type-I seesaw mechanism, as shown in Fig.~\ref{fig:type-i}.
The corresponding neutrino mass matrix, in the basis $(\overline{\nu}_{L_i}, \overline{N}_{L_{i'}})$ and $(\nu_{R_k}, N_{R_j})^T$, is given by
\begin{align} \label{eq:numat}
   \mathcal{M}_{\nu N}= \begin{pmatrix}
        0 & f v/\sqrt{2}\\
        g u/\sqrt{2} & M
    \end{pmatrix},
\end{align}
where $f$ and $M$ are $3 \times 3$ matrices and $g$ is $3 \times 2$ matrix with elements $f_{ij}$, $M_{i'j}$, $g_{i'k}$ respectively. 
For $M_{i'j} \gg f_{ij}v, g_{i'k}u$, the mass matrix in Eq.~\eqref{eq:numat} can be block diagonalized via the seesaw mechanism~\cite{Schechter:1981cv}. The resulting effective $3 \times 2$ light neutrino mass matrix realizes the Dirac type-I seesaw and is given by~\cite{CentellesChulia:2016rms,CentellesChulia:2018gwr}
\begin{align} \label{eq:numass}
    m_{\nu}= \frac{v u}{2} f M^{-1} g .
\end{align}
The matrix $m_{\nu}$ has rank two; consequently, only two neutrinos acquire masses while one remains massless, which is compatible with current neutrino oscillation data \cite{deSalas:2020pgw}.

We would like to mention that $\xi$ connects the exponential growth of dark matter with the neutrino sector. Its VEV not only drives the exponential dark matter growth but also determines the dark matter mass and contributes to neutrino mass generation. The  dark matter mass is given by (see App. \ref{sec:scalar} for details): 
\begin{align} \label{eq:dmmass}
    m^2_{\rm DM} \equiv  m^2_{\chi}=\mu^2_{\chi} + \frac{1}{2} \left(\lambda_{\Phi \chi} v^2 + \lambda_{\xi \chi} u^2 \right). 
\end{align}
Although Eqs. \eqref{eq:numass} and \eqref{eq:dmmass} show that both the neutrino and dark matter masses depend on the VEV of $\xi$, there is no direct one-to-one correspondence, since the dark matter mass also depends on other model parameters as well.
Nevertheless, such a connection between the dark matter and neutrino sectors is well motivated in a unified description and provides a simple, elegant solution to two of the major shortcomings of the SM. At the same time, the existence of a common underlying connection does not imply that the discovery of one phenomenon necessarily guarantees the observation of the other. Thus, the relation between the two sectors should be viewed as a theoretical connection within the model rather than as a direct one-to-one experimental correspondence.

\FloatBarrier
\section{Conclusion} 
\label{sec:conc}
The particle nature of dark matter and its production mechanism remain one of the long-standing unresolved problems in the field of particle and astroparticle physics. Broadly, there are two 
mechanisms -- thermal and non-thermal by which dark matter could be produced in the early universe. The former production mechanism usually requires larger couplings in comparison to the latter mechanism and hence can be probed easily at terrestrial experiments. However, so far null observations at various experiments constrain the thermal scenarios severely, making non-thermal scenarios more plausible. 

In this work, we focus on a particular non-thermal mechanism where the production of dark matter is 
thermally assisted. We study the exponential growth of dark matter, where the number density of dark matter grows exponentially 
before the population freezes. This is achieved in a simplified framework with a scalar dark matter candidate 
$\chi$ charged under $Z_3$ symmetry and a thermal bath particle $\xi$ which is a singlet. 
The interaction $\xi \chi^3$ then governs the exponential growth of dark matter. 

The exponential production mechanism is different from the freeze-in and freeze-out mechanisms. 
For a given mass of dark matter, the coupling required for getting the correct relic abundance lies between the freeze-in and freeze-out scenarios (see Fig.~\ref{fig:phasedigintro}). 
For IR freeze-in, the final abundance is independent of the dark matter mass, but for exponential and freeze-out scenarios, the abundance changes with the dark matter mass. Hence, an upper mass limit on both these scenarios can be determined. For the freeze-out, there exist an upper mass limit on the mass of the dark matter obtained from S-matrix unitarity and is around 100 TeV.

In this work, we determine the upper limit on the mass of the dark matter produced in the exponential growth using tree level perturbativity. 
The bounds are determined using a simplified model of a scalar dark matter $\chi$ interacting with the thermal bath particle $\xi$. In this scenario, 
the final dark matter abundance depends on four parameters -- initial dark matter abundance, dark matter mass, bath particle mass, and the coupling. We first determine the upper mass bound for the simplistic scenario where $m_\chi=m_\xi$. We note two observations: 
\begin{enumerate}
    \item For the same initial comoving abundance, i.e., $Y_\chi(x_{\rm init})$, increasing the dark matter mass increases the coupling required to establish the correct abundance. 
    \item For the same dark matter mass, decreasing the initial comoving abundance also increases the coupling required to establish the correct dark matter abundance. 
\end{enumerate}
Hence, from the arguments of tree level perturbativity, we cannot arbitrarily increase the dark matter mass and simultaneously decrease the initial abundance. At some scale, we hit essentially the perturbativity bound. This happens approximate at the scale of $\mathcal{O}(10^{15}) \ \mathrm{GeV}$. 
Beyond this scale, the required coupling to get the correct abundance is non-perturbative.  
This result differs significantly from the IR-freeze-in scenario, where strong bounds on the mass of the dark matter do not exist and the heavy dark matter also has feeble couplings. Thus, in exponential production of dark matter, it is possible to have a heavy dark matter with very large couplings, making it phenomenologically interesting. 
We also repeat the analysis for $m_\chi \neq m_\xi$ and find that for cases where the mass of the bath particle is lighter than the dark matter, the scattering becomes Boltzmann suppressed. Here the upper mass bounds on dark matter are smaller than in the case where $m_\xi = m_\chi$. Note that our analysis of finding the upper bound is model-dependent, as we use the criteria of perturbativity, where we essentially put perturbative requirements on the interaction strength between dark matter and the bath particle. The model dependence also enters on the ratio of the masses of the bath particle and the dark matter particle. Despite these variables,  our analysis and the upper bound will hold in any scenario where the renormalizable couplings driving the exponential growth effectively reduce to $\chi^3 \xi$ interactions.

We also extend our simplified framework to a UV-complete framework where we connect the dark matter sector to the neutrino sector. Here the accidental $U(1)_{B-L}$ symmetry and its residual $Z_3$ subgroup essentially connect exponential dark matter production with Dirac neutrino mass generation. More generally, residual $Z_{3n}$ symmetries can accommodate exponential dark matter growth while allowing either Dirac or Majorana neutrinos, depending on the symmetry realization. 

Thus, in our manuscript, we determined an upper mass bound of dark matter for the exponential growth mechanism using a simplified framework. This bound is interesting as it opens doors to heavy dark matter with $\mathcal{O}(1)$ couplings.

\section{Acknowledgments}
\noindent
DB would like to acknowledge Instituto de Fisica, Universidade de São Paulo, Brazil, where this work was started. She would also like to acknowledge the computational resources provided by IFUSP for this work. RK is supported by the National Research Foundation of Korea under Grant NRF-2023R1A2C100609111.
We would also like to thank the organizers of the WDMAP 2024
workshop, where this project was initiated.
\appendix

\section{Comparing exponential growth scenarios with other production mechanisms} \label{app:finout}
In this work, we focused extensively on exponentially produced dark matter through the interaction Lagrangian given in Eq.~(\ref{eqn:simp}), where dark matter 
is charged under $Z_3$ symmetry.  This symmetry, however, also allows for dark matter production by freeze-in, freeze-out, semi-annihilations, or decays wherever applicable. In our work, we considered the regions of the parameter space where the other processes were suppressed. In this appendix, we list out various domains where the other production mechanism can also contribute to the dark matter production and compare it with the exponential growth mechanism.

For $Z_3$ symmetric dark matter, the relevant terms of the Lagrangian are given by
\begin{eqnarray}
    \mathcal{L}_{\rm DM-int} = \frac{\lambda_{\rm exp}}{3!} \xi \chi^3 + {\lambda_{\rm fio}} \xi^\dagger \xi \chi^\dagger \chi  +  
    \frac{\lambda_{{\rm self}_1}}{3!} \chi^3 + 
    \frac{\lambda_{{\rm self}_2}}{2!} \left(\chi^\dagger \chi \right)^2 + \lambda_{\rm decay} \xi \chi^\dagger \chi + \text{h.c.}  \;.
    \label{eqn:intLag}
\end{eqnarray}
The Lagrangian in Eq.~\eqref{eqn:intLag} allows for dark matter 
production through various processes:  
freeze-out ($\chi^\dagger \chi \rightleftarrows \xi^\dagger \xi$), freeze-in ($\xi^\dagger  \xi \to \chi^\dagger \chi$), semi-annihilations ($\chi \xi \rightleftarrows \chi^\dagger \chi^\dagger$), exponential production ($\chi \xi \to \chi^\dagger \chi^\dagger$), three body decay ($\xi \to \chi^\dagger \chi^\dagger \chi^\dagger$), two body decay ($\xi \to \chi^\dagger \chi$) and $3 \to 2$ production ($\chi \chi \chi \rightleftarrows \chi^\dagger \chi$).  
Below we describe in detail the regions where different mechanisms can potentially dominate: 

\textbf{Freeze-out (2 $\to$ 2)}: This process is present when the dark matter population is in thermal equilibrium. For $2\to 2$ dark matter scattering, i.e., $\chi^\dagger \chi \rightleftarrows \xi^\dagger \xi$, to be the dominant production mechanism. 

\textbf{Freeze-out (3 $\to$ 2)}: For $3 \to 2$ freeze-out, dark matter is also in equilibrium with the thermal bath. However, in this case 
the interactions of dark matter are only enough to establish kinetic equilibrium with the thermal bath particles; 
the usual number-changing $ 2 \to 2$ processes are suppressed. The relic abundance in this case is essentially 
established by strong self-interactions through $3 \to 2$ processes. However, in the case of suppressed self-interactions, 
$3 \to 2$ scatterings do not result in the 
 dominant production of dark matter~\cite{Hochberg:2014dra}. 

\textbf{Freeze-in}: This mechanism works when the population of dark matter is not in thermal equilibrium in the early universe. 
In this case, the bath particles scatter to produce dark matter particles. The same interaction $\chi^\dagger \chi \xi^\dagger \xi$ which 
is responsible for dark matter production through freeze-out in the $2 \to 2$ case works for freeze-in. The only difference is that 
the coupling strength $\lambda_{\rm fio}$ required to get the correct relic abundance is feeble~\cite{Bernal:2017kxu}. The freeze-in 
production of dark matter can also happen via direct decay of the bath particles to dark matter particles if the process is kinematically allowed.

\textbf{Semi-annihilation}: This also lies in the category where dark matter is in thermal equilibrium in the early universe. 
Based on the symmetries, here dark matter annihilates a bit differently, i.e., $\chi \xi \to \chi^\dagger \chi^\dagger$~\cite{DEramo:2010keq}. This becomes the dominant 
mode when allowed by symmetries. Typically the coupling strength required to get the correct relic abundance is of the 
same order as $2 \to 2$ freeze-out, with notable exceptions in scenarios with CP violation~\cite{Ghosh:2020lma}. 

\textbf{Exponential production}: For this to be the dominant mechanism, $\lambda_{\rm fio} \ll \lambda_{\rm exp}$, and $\lambda_{\rm exp}$ must not be large enough to bring dark matter into equilibrium with the thermal bath. This dark matter production mechanism is complementary to semi-annihilation. Here, the dark matter is produced non-thermally, via the process $\chi \xi \to \chi^\dagger \chi^\dagger$, as this is complementary to semi-annihilation.

We primarily focus on the production 
of dark matter through scattering processes with the bath particle --- hence we neglect the dark matter self-scatterings and decays of the 
bath particle to the dark matter particle. We work in the kinematic limit where $m_\xi < 2 m_\chi$ to avoid both three and two body decays. 
With these assumptions, we are left with the following terms in the interaction Lagrangian --- $\frac{\lambda_{\rm exp}}{3!} \xi \chi^3 + \frac{\lambda_{\rm fio}}{2!} \xi^\dagger \xi \chi^\dagger \chi $. Depending on the coupling strengths, we can have dark matter production by either freeze-in, freeze-out, semi-annihilation, or exponential 
production mechanisms.  
The relic density of dark matter in this subset scenario can be determined by solving the zeroth-moment Boltzmann equation
\begin{align} \label{eq:density}
    \Dot{n}_{\chi} + 3 H n_{\chi} = \langle \sigma v\rangle_{\rm exp} \left( n_{\chi} n_{\xi} - \frac{n_\xi}{n_\chi^{\rm eq}} n_\chi^2 \right) + \langle \sigma v\rangle_{\rm fio} \left( {(n^{\rm eq}_{\chi})^2} - n^2_{\chi} \right),
\end{align}
where $H$ is the Hubble rate, $\langle \sigma v\rangle_{\rm exp}$ $\left(\langle \sigma v\rangle_{\rm fio}\right)$ 
is the thermally averaged cross section for the process $\chi \xi \to \chi \chi$ ($\chi \chi \to \chi \xi$), 
and $n_{\xi}$ is the number density of the bath particle $\xi$ in equilibrium. Note that in writing this equation, we 
have implicitly assumed that dark matter at least traces some scaled version of the thermal distribution function as in Eq.~(\ref{eqn:scaled}).
Rewriting the equation in terms of the comoving number density, we obtain
\begin{eqnarray}
   \frac{d Y(x)}{d x} &=&  \frac{h_{\rm eff}(x) s(x)}{x H(x)} \bigg[ \langle \sigma v\rangle_{\rm exp}  \left(Y_\chi Y_\phi^{\rm eq}  - Y_\chi^2\right) +    \langle \sigma v\rangle_{\rm fio} \left({Y^{\rm eq}_\chi}^2 - Y_\chi^2\right)   \bigg]\;.
 \label{eqn:nodensity}
\end{eqnarray}

\begin{figure}[h!]
    \centering
    \includegraphics[width=0.6\textwidth]{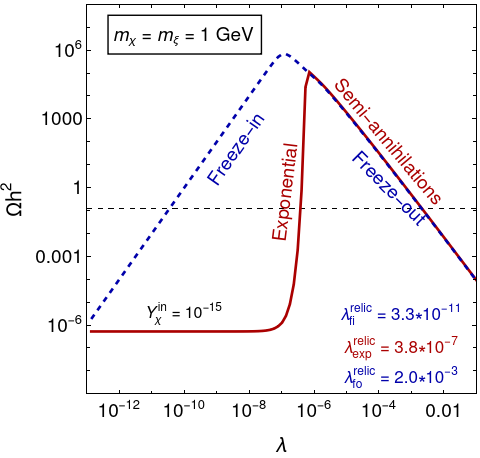}   
    \caption{The phase diagram for various dark matter production mechanisms is shown for $m_\xi =m_{\rm DM} = 1$ GeV. The complementary parameter regions governed by exponential production and semi-annihilation are shown in red. Similarly, the freeze-in and freeze-out regimes are depicted with dashed blue curves. The horizontal black dashed line indicates the relic abundance of dark matter consistent with observations~\cite{Planck:2018vyg}. Here, the coupling $\lambda$ represents $\lambda_{\rm exp}$ for exponential and semi-annihilation, and $\lambda_{\rm fi}$ for freeze-in and freeze-out.}
    \label{fig:phasedig1}
\end{figure}

We solve the Boltzmann equation numerically in {\tt{Mathematica}} software \cite{Mathematica} as a function of $m_\chi$, $m_\xi$, $\lambda_{\rm exp}$, $\lambda_{\rm fio}$ and initial abundance of dark matter $Y_\chi({x_{\rm init}})$. 
The relic abundance can then be simply determined, assuming no entropy dilution, as
\begin{align} \label{eq:relic}
    \Omega h^2 \equiv \frac{\rho_{\rm DM}}{\rho_c/h^2}= \frac{m_{\rm DM} n_\chi}{\rho_c/h^2} = \frac{m_{\rm DM} s_{\rm today} Y_\chi(x_{\rm today})}{\rho_c/h^2},
\end{align}
where $\rho_c$ is the critical density and $h$ is the dimensionless Hubble parameter.
We illustrate the qualitative features of dark matter production and its comparisons with other dark matter production mechanisms through the phase diagram presented in Fig.~\ref{fig:phasedig1}. 
For the plot, we consider that either $\lambda_{\rm exp}$ is present or $\lambda_{\rm fio}$ is present at a time. This helps us to bring out the contrasting features between freeze-in and exponential production and, at the same time, helps to bring out the complementary features between exponential production and semi-annihilations as well as complementarity between freeze-in and freeze-out. 
In order to focus on the key features of the mechanism, we fix $m_{\xi} = m_{\rm DM} =1$ GeV for the dark matter particle ($\chi$) and the bath particle ($\xi$). In the figure, we show dark matter abundance as a function of coupling ($\lambda$). We choose neutral label $\lambda$ which can indicate either $\lambda_{\rm exp}$ or $\lambda_{\rm fio}$. 
The various production mechanisms\footnote{The phase diagrams presented in Fig.~\ref{fig:phasedig1} correspond to either $\lambda= \lambda_{\rm exp} \ (\text{and} \  \lambda_{\rm fi}=0)$ shown in red or $\lambda= \lambda_{\rm fi} \ (\text{and} \ \lambda_{\rm exp}=0)$ shown in blue dashed curves. } are depicted in Fig.~\ref{fig:phasedig1}. The regions corresponding to exponential production and semi-annihilation are shown in red, emphasizing their complementary roles in shaping the dark matter relic abundance. For comparison, the conventional freeze-in and freeze-out regimes, which are also complementary to each other, are indicated by dashed blue curves. This phase diagram provides a unified view of how the various production mechanisms populate distinct regions of parameter space and determine the relic density in their respective regimes.

 From Fig.~\ref{fig:phasedig1}, it is evident that the transition from exponential production to semi-annihilation is sharp, exhibiting exponential features, 
 whereas the crossover between freeze-in and freeze-out remains comparatively smooth. 
 This behavior can be understood from the distinct dependence of the dark matter abundance on the coupling $\lambda$. 
 In the exponential production regime, the abundance grows as $e^{\lambda^2}$, leading to a rapid enhancement, while in the semi-annihilation regime it decreases as $1/\lambda^2$. In contrast, freeze-in exhibits $\lambda^2$, and freeze-out follows a $1/\lambda^2$ dependence, resulting in a more gradual transition between these regimes. The observed dark matter relic abundance crucially depends on the mode of production and the corresponding coupling $\lambda$. The relic density is therefore determined by the $\lambda$-scaling specific to each production regime.
Hence, exponential production and freeze-in are active in the small and feeble coupling regions, respectively, while their complementary parts, semi-annihilation and freeze-out, take over as the coupling becomes large.

\section{Scalar potential and mass spectrum} \label{sec:scalar}
The scalar sector of the model consists of three fields: the Higgs doublet $\Phi$, the singlet scalar $\xi$, and the scalar dark matter candidate $\chi$. The corresponding invariant scalar potential, consistent with the charge assignments in Tab.~\ref{tab:particle}, is given by\footnote{Here, the \UBL \ symmetry is considered global, giving rise to a massless Goldstone boson. The inclusion of soft symmetry breaking terms in the scalar potential can lift its mass. However, in the gauged version of the model, the corresponding Goldstone boson will be absorbed by the $Z'$ boson.}
\begin{align} \label{eq:scalpot}
    V&= -\mu_{\Phi}^2 \Phi^\dagger \Phi - \mu_{\xi}^2 \xi^\dagger \xi  + \mu_{\chi}^2 \chi^\dagger \chi +  \frac{\lambda_{\Phi}}{2} \left(\Phi^\dagger \Phi \right)^2 +  \frac{\lambda_{\xi}}{2} \left(\xi^\dagger \xi \right)^2 +  \frac{\lambda_{\chi}}{2} \left(\chi^\dagger \chi \right)^2  \nonumber \\ 
    & +  \lambda_{\Phi \xi}\left(\Phi^\dagger \Phi \right)\left(\xi^\dagger \xi \right) + \lambda_{\Phi \chi} \left(\Phi^\dagger \Phi \right)\left(\chi^\dagger \chi \right) 
    +\lambda_{ \xi \chi} \left(\xi^\dagger \xi \right)\left(\chi^\dagger \chi \right)+ \frac{\lambda'_{\xi \chi}}{3!} \left(\xi^\dagger \chi \chi \chi  + \text{H.c.} \right).
\end{align}
The last term in Eq.~\eqref{eq:scalpot} serves as a cornerstone of our model, as it governs the exponential production of dark matter candidate $\chi$. In particular, this interaction facilitates the conversion of the thermal bath particle $\xi$ to dark matter candidate $\chi$.
Its presence is not \textit{ad hoc}; rather, it is allowed by the residual $Z_3$ symmetry that emerges from the spontaneous breaking of the $U(1)_{B-L}$ \ symmetry. Notably, the same residual $Z_3$ symmetry also ensures that neutrinos remain Dirac particles \cite{Bonilla:2018ynb}. 

We now proceed to analyze the scalar mass spectrum. After symmetry breaking, the Higgs doublet $\Phi$ and singlet scalar $\xi$ acquire vevs. In the unitary gauge, $\Phi$ and $\xi$ can be expanded as follows
\begin{align} \label{eq:exp}
\Phi &=\frac{1}{\sqrt{2}} 
\begin{pmatrix}
0\\
v+h'
\end{pmatrix},\,  \quad \xi= \frac{1}{\sqrt{2}}(u + \xi' ) .
\end{align}
The conditions corresponding to the tadpole equations are obtained by minimizing the scalar potential, and are explicitly given by
\begin{align}
    \mu^2_{\Phi}&=\frac{1}{2}\left(\lambda_{\Phi} v^2 + \lambda_{\Phi \xi} u^2 \right), \nonumber \\
     \mu^2_{\xi}&=\frac{1}{2}\left(\lambda_{\xi} u^2 + \lambda_{\Phi \xi} v^2 \right).
\end{align}
The fields $h'$  and $\xi'$ mix with each other, and their mass-squared matrix in the $(h', \xi')$ basis is given as follows
\begin{align}
    \mathcal{M}^2_{h' \xi'}=\begin{pmatrix}
\lambda_{\Phi} v^2 & \lambda_{\Phi \xi} u v\\
\lambda_{\Phi \xi} u v & \lambda_{\xi} u^2 
\end{pmatrix}.
\end{align}
Let us denote the physical scalar mass eigenstates by $\boldsymbol{h}$ and $\boldsymbol{\xi}$, which arise from the mixing of the interaction eigenstates through an angle $\theta$. Accordingly, the mass eigenstates can be expressed in terms of the interaction eigenstates through the following relation
\begin{align}
    \begin{pmatrix}
        \boldsymbol{h} \\
        \boldsymbol{\xi}
    \end{pmatrix}=
    \begin{pmatrix}
        \cos{\theta} & -\sin{\theta} \\
        \sin{\theta} & \cos{\theta}
    \end{pmatrix}
     \begin{pmatrix}
        h' \\
         \xi'
    \end{pmatrix} ,
\end{align}
and the mixing angle $\theta$ is given by 
\begin{align}
    \tan{2\theta}= \frac{2 \lambda_{\Phi \xi} u v}{ \lambda_{\xi} u^2-\lambda_{\Phi} v^2} \ .
\end{align}
The masses of physical scalars $\boldsymbol{h}$ and $\boldsymbol{\xi}$ can be computed as follows
\begin{align}
    m^2_{\boldsymbol{h}}&=\frac{1}{2} \left( \lambda_{\Phi} v^2+\lambda_{\xi} u^2 -\sqrt{\left(\lambda_{\xi} u^2-\lambda_{\Phi} v^2 \right)^2 +4 \left( \lambda_{\Phi \xi} u v\right)^2 } \right), \\
    m^2_{\boldsymbol{\xi}}&=\frac{1}{2} \left( \lambda_{\Phi} v^2+\lambda_{\xi} u^2 +\sqrt{\left(\lambda_{\xi} u^2-\lambda_{\Phi} v^2 \right)^2 +4 \left( \lambda_{\Phi \xi} u v\right)^2 } \right).
    \label{eqn:mxi}
\end{align}
We identify $\boldsymbol{h}$ as the 125 GeV Higgs boson. The mass of the dark matter candidate $\chi$ is given by
\begin{align}
    m^2_{\rm DM} \equiv  m^2_{\chi}=\mu^2_{\chi} + \frac{1}{2} \left(\lambda_{\Phi \chi} v^2 + \lambda_{\xi \chi} u^2 \right). 
\end{align}
%

\bibliographystyle{utphys}
\bibliography{references} 
\end{document}